\documentclass{PoS}

\newcommand{\HG}{\hat{\Gamma}}
\newcommand{\be}{\begin{equation}}
\newcommand{\ee}{\end{equation}}
\newcommand{\ba}{\begin{array}}
\newcommand{\ea}{\end{array}}
\newcommand{\baa}{\begin{array}}
\newcommand{\eaa}{\end{array}}
\newcommand{\bea}{\begin{eqnarray}}
\newcommand{\eea}{\end{eqnarray}}

\newcommand{\kb}{\bar{k}} 
\newcommand{\pc}{p^{(c)}} 
\newcommand{\ps}{p^{(s)}} 

\newcommand{\ET}{{\cal E}}

\newcommand{\ev}{\hat{e}}

\newcommand{\N}{L}
\newcommand{\lef}{l^{\rm eff}}
\newcommand{\ttheta}{\tilde \theta}

\title{Perturbative analysis of twisted volume reduced theories}

\ShortTitle{Perturbative analysis of twisted volume reduced theories}

\author{\speaker{Margarita Garc\'{\i}a P\'erez}%
         \\
        Instituto de F\'{\i}sica Te\'orica UAM/CSIC, Universidad Aut\'onoma de Madrid, 
        E-28048-Madrid, Spain\\
        E-mail: \email{margarita.garcia@uam.es}}

\author{Antonio Gonz\'alez-Arroyo\\
        Instituto de F\'{\i}sica Te\'orica UAM/CSIC and  Departamento de F\'{\i}sica Te\'orica, C-15,
        Universidad Aut\'onoma de Madrid, E-28049-Madrid, Spain\\
        E-mail: \email{antonio.gonzalez-arroyo@uam.es}}
\author{Masanori Okawa\\
        Graduate School of Science, Hiroshima University,
        Higashi-Hiroshima, Hiroshima 739-8526, Japan\\
        E-mail: \email{okawa@sci.hiroshima-u.ac.jp}}

\abstract{
We discuss the perturbative expansion of  SU(N) Yang-Mills
theories defined
on a d-dimensional torus of linear size $l$ with twisted boundary
conditions,
generalizing previous results in the literature.
For a specific class of twist tensors depending on a single integer
flux value $k$, we show that perturbative results to all orders depend
on the combination $l N^{2/d}$  and a flux-dependent angle $\tilde
\theta$.
This implies a new kind  of volume independence that holds at finite N
and for fixed values of $\tilde \theta$. Our results also provide
interesting information about the possible occurrence of tachyonic
instabilities at one-loop order. We  support the prescription
that instabilities are avoided, if the large N limit is taken keeping
$\tilde \theta >\tilde \theta_c$, and appropriately scaling the magnetic flux $k$ with $N$. 
Numerical results in 2+1 dimensions provide a
test of  how these ideas extend into the non-perturbative regime.}

\FullConference{31st International Symposium on Lattice Field Theory LATTICE 2013\\
                 July 29 - August 3, 2013\\
                 Mainz, Germany}

\begin{document}

\section{Introduction}

In a recent paper \cite{Perez:2013dra}, focused on the study of 2+1 dimensional SU($N$) Yang-Mills theory
defined on a 2-d torus, we have analyzed the interplay between the rank of the group $N$ and the finite 
volume effects in the presence of a non-trivial magnetic flux. We have presented perturbative and numerical 
evidence of a kind of volume independence in the theory,  reflected in the combined $Nl$
dependence of physical quantities, $l$ being the size of the 2-dimensional spatial manifold. 
Here, we will generalize the perturbative results to the case of a four dimensional set-up with the 
kind of twisted boundary conditions relevant for Twisted Eguchi-Kawai (TEK) reduction, as put forward by 
two of the present authors \cite{TEK1,TEK2}. We will start by presenting the perturbative set-up for a 
Yang-Mills theory with twisted boundary conditions, and the interplay between finite volume and 
finite $N$ effects in this context.
We will continue by discussing possible caveats to the volume independence conjecture, 
including the presence of tachyonic instabilities \cite{Guralnik:2002ru}
and the appearance of symmetry breaking in the TEK model for large values of $N$ \cite{IO}. 
We will argue that the
prescription to scale the magnetic twist with $N$, put forward in Refs. \cite{TEK1,TEK2} to prevent 
the latter, also succeeds in avoiding the tachyonic behaviour. We will end by presenting
some numerical results in 2+1 dimensions that provide a test on the realization of volume independence 
at a non-perturbative level.

\section{Perturbative set-up}
To set the stage, let us start with a brief and general introduction on the formulation of the 
SU(N) Yang-Mills theory with twisted boundary conditions. The notation and the discussion will follow 
the review \cite{TONYREV}. We start with a manifold composed of a d-dimensional torus, of lengths $l_\mu$, 
times some non-compact extended directions. The latter will not be relevant for our purposes and 
will be neglected in the discussion below. Gauge fields 
on this base space satisfy periodicity conditions in the compact directions given by:
\be
\label{A_per_cond}
A_\mu(x+l_\nu \ev_\nu)= \Omega_\nu(x) A_\mu(x)\Omega^\dagger_\nu(x)+
i \Omega_\nu(x)
\partial_\mu\Omega^\dagger_\nu(x)\, ,
\ee
where the SU($N$) twist matrices $\Omega_\mu (x)$ are subject to the consistency conditions: 
\be
\Omega_\mu(x+l_\nu \ev_\nu)\Omega_\nu(x)= Z_{\mu\nu}  \Omega_\nu(x+l_\mu \ev_\mu)\Omega_\mu(x)\, ,
\ee
with $Z_{\mu\nu} = \exp \{2 \pi i n_{\mu \nu} / N\}$, and $n_{\mu \nu}\in Z\!\!\!Z_N$. 
In what follows, we will focus on the case of constant twist matrices $\Omega_\mu(x)=\Gamma_\mu$.
They are known under the name of twist-eaters. For the so-called irreducible twists,
it can be shown that they are uniquely defined modulo global gauge transformations 
(similarity transformations) and  multiplication by an element of $Z\!\!\!Z_N$
\cite{TONYREV}. We will be considering here the case of even number of twisted compactified directions $d$.  
If we choose $N= L^{d/2}$, with  $L \in Z\!\!\!Z$, the twist  $n_{\mu \nu} = \epsilon_{\mu \nu} \, k \, N/L$, 
with $\epsilon_{\mu \nu} = \Theta(\nu-\mu)- \Theta(\mu-\nu)$ (where $\Theta$ is the step function), 
is irreducible if $k$ and $L$ are co-prime. In what follows we will focus on this specific twist choice.

The periodicity constraint:
\be
A_\mu(x+ l_\nu \, \hat\nu)= \Gamma_\nu A_\mu(x)\Gamma^\dagger_\nu \ ,
\ee
can be resolved by introducing a basis of the space of $N\times N$ matrices satisfying:
\be
\Gamma_\mu\HG(\pc)\Gamma_\mu^\dagger = e^{i l_\mu \pc_\mu} \HG(\pc) \quad .
\ee
The basis can be constructed in terms of the twist matrices as:
\be
\HG(\pc) = \frac{1}{\sqrt{2N}}\, e^{i \alpha(\pc)} \Gamma_0^{s_0} \cdots \Gamma_{d-1}^{s_{d-1}}
\ee
with $\alpha(\pc)$ an arbitrary phase factor, and: 
\be
\pc_\mu = \frac{2 \pi }{\N \, l_\mu } \, \epsilon_{\mu\nu} k\, s_\nu \, ,
\ee
with $s_\nu$ integers defined modulo $\N$. In the case of irreducible twists, 
one can show that there are $N^2$ linearly independent such
matrices which, excluding the identity, provide a basis of the $SU(N)$ Lie algebra.   
The phase factors, $\alpha(\pc)$, can be chosen to satisfy the following
commutation relations:
\be
[\hat \Gamma(p), \hat \Gamma (q)] = i \, F(p, q , -p-q)) \, \hat \Gamma(p+q) \, ,
\ee
with  
\be
F(p,q,-p-q)= -\sqrt{\frac{2}{N}}  \, \sin\left(\frac{ \theta_{\mu \nu}}{2} \, p_\mu q_\nu
\right) \, ,
\ee
playing the role of the SU(N) structure constants in this particular basis.
Here, 
\be
\label{theta}
\theta_{\mu \nu} =    \frac{ \N^2  \, l_\mu \,  l_\nu} {4\pi^2} \times  \, \tilde
\epsilon_{\mu \nu} \, \tilde \theta  \, ,
\ee
with $\tilde \epsilon_{\mu \nu} \epsilon_{\nu\sigma} = \delta_{\mu \sigma}$,
and the angle $\tilde \theta = 2 \pi \bar k / \N$, 
where $\bar k$ is an integer satisfying $k \bar k = 1$ (mod $\N$).

We can now expand the gauge fields in this basis:
\be
A_\nu(x)=  {1 \over \sqrt{V}}\, \sum'_{p}   e^{i p \cdot x}\, \hat{A}_\nu(p)\, \hat \Gamma(p) \ ,
\ee
with momenta decomposed as $p_\mu =  \ps_\mu + \pc_\mu$,  the sum of a colour-momentum 
part $\pc_\mu$, and a spatial-momentum part quantized in the usual way: $\ps_\mu=2 \pi m_\mu/l_\mu$, 
with $m_\mu \in Z\!\!\!Z$.
The prime implies that the zero colour-momentum  component is excluded from the sum.
Neglecting this issue, the momentum is quantized as if the theory lived in an effective torus
with sizes $\lef_\mu = \N \, l_\mu $. 
Note also that the Fourier coefficients $\hat{A}_\nu(p)$ are just complex numbers, 
so that all the effect of the colour is translated into the momentum dependence of the 
group structure constants.

One can now easily generalize the Feynman rules derived for $d=2$ in Ref.~\cite{Perez:2013dra},
to arrive to the conclusion that all vertices in perturbation theory are proportional to the factor:
\be
{g \over \sqrt{V}} F(p,q,-p-q) = - \sqrt{\frac{2\lambda} {V_{\rm eff}}}\, 
\, \sin\Big (\frac{\theta_{\mu \nu}}{2} \, p_\mu  q_\nu\Big ) \, , 
\ee
where $V_{\rm eff} \equiv \prod_\mu \lef_\mu$. 
This peculiar momentum dependent Feynman rules relate the twisted theory with 
a non-commutative Yang-Mills theory with non-commutativity parameter $\theta_{\mu \nu}$ 
\cite{GAKA}-\cite{ambjorn}.

\section{Volume independence}

Let us now analyze the dependence of the results on the rank of the gauge group $N$ and the 
sizes of the torus $l_\mu$.  Recalling the definition of $\theta_{\mu \nu}$ in 
Eq.~(\ref{theta}) and the momentum quantization rule, it is clear that all the $N$ and $l_\mu$ 
dependence enters only through the combination  $\lef_\mu = \N l_\mu$, and the 
angle $\ttheta = 2 \pi \bar k / \N$. This implies that the perturbative expansion, 
at fixed $\ttheta$ and $\lef$,  
depends in an indistinguishable way on $N$ and the torus size.
We dub this phenomenon {\it volume reduction} or {\it volume independence} at finite $N$.
A limiting case would be TEK reduction which applies to a discretized version
of the Yang-Mills theory in which $l_\mu=a$ (the lattice spacing). 
As a matter of fact, TEK models have been used as a regularized version of non-commutative gauge theories with
non-commutativity parameter:
\be
\theta_{\mu \nu}^{\rm TEK} =    \frac{ \N^2  \, a^2} {4\pi^2} \times  \, \tilde
\epsilon_{\mu \nu} \, \tilde \theta  \, ,
\ee

Beyond perturbation theory, there are, however, possible caveats to the volume independence conjecture.  
As mentioned previously, several authors realized  the presence of 
tachyonic instabilities in certain non-commutative 
theories~\cite{Guralnik:2002ru}. These 
extend to ordinary theories with twisted boundary conditions and
present a menace to the volume independence mechanism. The problem 
occurs at one loop in perturbation theory. In Ref.~\cite{Perez:2013dra} 
we saw how the
problem does not arise in 2+1 dimensions if one adopts the large N 
prescription given in  Ref.~\cite{TEK2}. The argument extends to 4 dimensions 
as well, and goes as follows. The transverse part of the 2-point vertex 
function has a non-zero value at leading order, since twisted boundary
conditions eliminate  zero-momentum gluons.  This contribution is 
proportional to $|p|^2\sim 4\pi^2/l_{\rm eff}^2$. At one loop, 
the self-energy  contribution is negative and proportional to 
\begin{equation}
\label{selfenergyNC}
\lambda \frac{\tilde{p}_\mu \tilde{p}_\nu}{|\tilde{p}|^4}
\end{equation}
where $\tilde{p}_\mu=\theta_{\mu \nu} p_\nu$. Instability arises 
if the second term is larger than the first. This occurs for large
enough $\lambda$ where the calculation is unreliable. However, since 
the first term goes to zero as $l_{\rm eff}$ goes to infinity, one might
wonder if instability could arise in that limit. The prescription given 
in Ref.~\cite{TEK2} amounts to taking the $l_{\rm eff}\longrightarrow \infty$
limit with $\theta_{\mu \nu}$ given by Eq.~(2.9) and keeping $\tilde{\theta}$
fixed. Plugging this expression in Eq.~(\ref{selfenergyNC}), one sees
that the negative self-energy also goes to zero as $l_{\rm eff}$ goes
to infinity and the critical $\lambda$ remains finite, and of order
$\tilde{\theta}^2$. Thus, as supported by our numerical results, no 
instability should arise for $\tilde{\theta}>\tilde{\theta}_c$. 
The previous evidence for instability occurred when taking a different
limit, in which  $\tilde{\theta}$ was decreasing to zero as $1/L$.

Additional complications could arise from the fact that it is impossible to 
strictly keep $\ttheta$ fixed as $N$ changes. 
This is so because $\ttheta$ is a rational number with coprime rational factors
$\bar k$ and $L$. Smoothness of physical quantities on $\ttheta$ is thus 
required for volume independence to hold. Our results for the electric flux 
energies and the perturbative glueball spectrum in 2+1 dimensions exhibit such 
a smooth dependence, but the issue is difficult to settle in general terms 
and it has indeed been discussed profusely in the context of non-commutative 
gauge theories without conclusive results 
(see e.g. \cite{Guralnik:2001pv}, \cite{AlvarezGaume:2001tv}).
Finally, effects arising from non-perturbative physics might not respect the $\lef$ dependence. There are indications that this is so for certain twist choices in TEK models, 
where reduction fails due to spontaneous symmetry breaking at large $N$ \cite{IO}.
It has been recently shown, however, that symmetry breaking can be 
avoided if (in addition to keeping $\ttheta> \ttheta_c$) the magnetic flux  
is scaled  with $L$, as $L$ goes to infinity \cite{TEK1,TEK2}.

Most of these questions can only be addressed non-perturbatively. In what follows we will summarize the 
results of a numerical analysis for the case of 2+1 dimensions \cite{Perez:2013dra}. 
The analysis for $d=4$ is ongoing and will be presented elsewhere.

\section{Non-perturbative results in 2+1 dimensional SU(N) Yang-Mills theory}

Our analysis will be focused on the study of the electric flux energies, ${\cal E}$,
extracted from  Polyakov loop correlators. 
It is easy to show that these operators carry electric flux $e_i= - \epsilon_{i j} n_j \bar{k}$,
determined by the gluon colour momentum $\vec p = 2 \pi \vec n / (Nl)$.
In perturbation theory at one-loop, a compact formula, exhibiting a smooth $\ttheta$
dependence, has been derived in Ref. 
\cite{Perez:2013dra}:
\be
\label{enerp}
{{\cal E}^2 \over \lambda^2}=   {  |\vec n|^2 \over 4 x^2 }
- { 1 \over x } G\Big ({\vec e \over N}\Big ) \, ,
\ee
where
\be
G  (z) = { 1 \over 16 \pi^2 } \int_0^{\infty} {dt \over \sqrt{ t}}
\Big(\theta_3^2(0,t) - \prod_{i=1}^2\theta_3(z,i t)- {1 \over t}\Big)\, ,
\ee
in terms of the Jacobi theta function: $
\theta_3(z,it) = \sum_{n \in {\bf Z}} \exp\{-t \pi n^2 + 2 \pi i n z\}$ .
We have introduced the dimensionless parameter $x = N l \lambda/ 4\pi $. Note that in 
3 dimensions the coupling constant is dimensionful. Thus, all energy scales can be expressed in
units of $\lambda$ and the resulting dimensionless quantities should appear in perturbation theory
as a power series in $\lambda l$. Combining this information with 
volume independence, we conclude that the relevant scale parameter in perturbation theory is 
precisely $x$  (for a similar statement involving $Nl \Lambda_{QCD} $ in 4-d see 
\cite{Unsal:2010qh} and  \cite{Guralnik:2001pv}).

\begin{figure}
\hspace*{1cm}\includegraphics[width=0.6\textwidth,angle=-90]{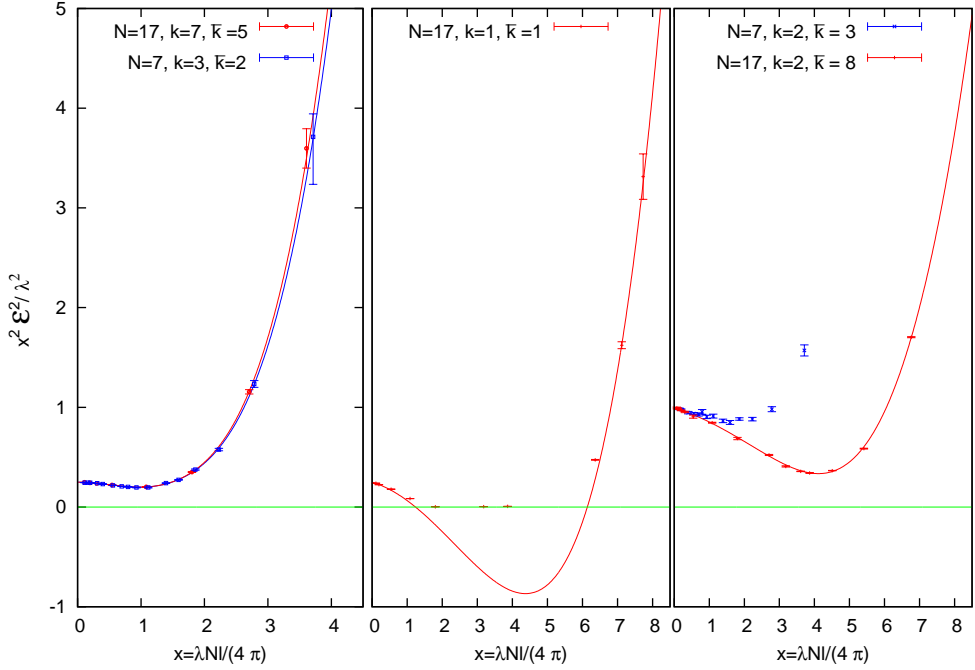}
\caption{We display $x^2 \ET^2 /\lambda^2$, as a function  of  $x$, for 
electric flux states with momentum: $p = (2\pi/Nl,0)$ (Left, Center), and $p = (4\pi/Nl,0)$ 
(Right).  }
\label{results}
\end{figure}

In Ref. \cite{Perez:2013dra} we have presented evidence that $x$-scaling holds beyond perturbation theory
for choices of the twist that do not exhibit tachyonic instability.
As an illustration,  we present in the left plot of Fig.~\ref{results} an analysis of electric-flux 
states with minimal momentum $\vec p =(2\pi/Nl, 0)$. We display the $x$ dependence of the combination 
$x^2 {\cal E}^2 /\lambda^2$, for $(N, \kb)=(7,2)$ and $(17,5)$, corresponding to very close values of $\ttheta$.
We stress the striking similarity of the results despite the very different values of $N$. Other results, 
for varying $N$ and $\kb$, confirm this conclusion, giving support to the conjecture of a
universal $x$-dependence for fixed $\ttheta$. 
The data show that the small-$x$ behaviour follows the perturbative formula, starting at the tree-level result
$x^2 {\cal E}^2 /\lambda^2 =0.25$.  At large torus sizes, we expect a linear growth of the  
energy that can be cast in a form that also exhibits $x$-scaling: 
\be
\label{conf}
{{\cal E} (\vec e/N) \over \lambda}  = 4 \pi x \, \frac {\sigma' }{\lambda^2} \, \phi\Big ({\vec e \over N} \Big ) \, ,
\ee
where the string tension, for electric flux $\vec e$, has been parametrized as: 
$\sigma_{\vec e} = N \sigma' \phi(\vec e /N)$. 
Higher order string corrections to this formula can also be taken into account, including 
the contribution of Kalb-Ramond $B$-fields which play an important role in the twisted  
set-up ~\cite{Guralnik:2001pv}.  Indeed, the observation that the $B$-field contribution amounts precisely 
to the perturbative tree-level term in Eq.~(\ref{enerp}) has guided us in the search for 
an $x$-dependent parameterization that fits very well the data.  The reader is referred to 
Ref.~\cite{Perez:2013dra} for further details and a full account of the results.
Incidentally, let us mention that we have analyzed the dependence of the string tension on the electric flux,
finding a clear preference for Sine scaling with $\phi(z) = \sin(\pi z)/\pi $, over Casimir scaling  with 
$\phi(z) = z(1-z)$.

Our results allow to analyze in detail the issue of tachyonic instabilities. The general perturbative argument 
presented in Sec. 3 can be refined using Eq.~(\ref{enerp}). 
As already discussed, the one-loop correction, being  negative, could give rise to a tachyonic excitation. 
This occurs above a critical 
coupling: $x_c(\vec{e})=|\vec n|^2 /(4 G(\vec{e}/N))$. The quantity $G(\vec z)$ diverges as $1/|\vec z|$  
for small $|\vec z|$. This would seem to unavoidably drive  $x_c$ to zero as $N$ goes to infinity. 
However, given the relation $n_i = - \epsilon_{ij} k e_j$, 
it suffices to scale $k \sim \sqrt{N}$ to 
push $x_c$ into the non-perturbative domain. 
On the opposite side, if we look at minimal momentum $|\vec n|=1$, the electric flux is given by $\bar k$
and the critical coupling occurs at $x_c(\vec{e})=4 \pi^2 \bar k /N\equiv 2\pi \tilde \theta$.  
Thus, keeping $\tilde \theta > \tilde \theta_c$ the perturbative instability is avoided.  

One can still worry about the possible appearance of non-perturbative instability. An argument based 
on the effective string description has been used in Ref.~\cite{Perez:2013dra}
to indicate that this is avoided if: 
$|\vec n| \, |\vec e| (N-|\vec e|) / N^2 > 1/12$. Still, a full proof should rely on numerical results. In 
our previous work~\cite{Perez:2013dra}, we have performed simulations at several values 
of $\ttheta$ and $k/N$. All the results with $\tilde \theta > 2 \pi /7$ and $k/N > 2/17$ showed no indication
of instability.  
A representative sample is presented in Fig.~\ref{results}.  
In addition to the stable case already discussed, we display two cases in which
$k/N$ becomes small. They correspond to electric flux $|\vec e |=1$, with  $k=\bar k=1$ (central plot)  and  $k=2$, $\bar k=(N-1)/2$ (right plot). 
In the first case (with $N=17$), as the energy squared decreases from the 
perturbative 
tree-level value  it touches, at some point, zero. This signals the appearance 
of an instability.  Of course, 
the energy does never become tachyonic and what we observe instead 
is a region where the electric 
flux has a small mass, possibly reflecting a non zero vacuum expectation value of the Polyakov loop. 
In the large $x$ regime though, the linearly 
rising potential overcomes this behaviour and restores the 
standard $x$-dependence.
In the second case, we display two values of  $k/N=2/7$ and $2/17$. They  are
still large enough to prevent the appearance of instability. However, 
we observe a strong decrease in the energy of electric flux with 
decreasing $k/N$. Assuming 
this trend continues, one expects an instability to  set-in below a given 
value of this ratio.

\section{Conclusions}

We have given a  unified description of the perturbative expansion
of SU(N) Yang-Mills theory on an even dimensional torus traversed
by  Z(N) magnetic flux through each plane. We stress the emergence
of an effective size parameter combining spatial and group degrees of
freedom. Our results, valid at finite $N$, provide important information
on large N reduced models and the twisted volume reduction mechanism. In
particular, they support the prescription given in Ref.~\cite{TEK2} on how the
flux has to scale with the rank N.  A few numerical results in 2+1
dimensions for the  energies of electric flux sectors support the
applicability of these ideas at the non-perturbative level. A good
description is given of the evolution of these energies at all scales.

\acknowledgments
M.G.P. and A.G-A acknowledge support from the grants FPA2012-31686 and 
FPA2012-31880,  the MINECO 
Centro de Excelencia Severo Ochoa Program SEV-2012-0249,
the Comunidad Aut\'onoma de Madrid  HEPHACOS
S2009/ESP-1473, and the EU  PITN-GA-2009-238353 (STRONGnet). They participate 
in the Consolider-Ingenio 2010 CPAN (CSD2007-00042).  
M. O. is supported by the Japanese MEXT grant No 23540310. We acknowledge the use of 
the IFT clusters.

\end{document}